\documentclass[11pt,a4paper]{article}
\usepackage{amsmath}
\usepackage[mathcal]{eucal}
\usepackage{latexsym}
\usepackage{amssymb}
\begin{titlepage}
\title{Instanton representation of Plebanski gravity.  Euclidean signature minisuperspace solution}
\author{Eyo Eyo Ita III}
\medskip
\input amssym.def
\input amssym.tex

\def \in{\indent}

\begin{document}
\maketitle
\bigskip
\centerline{Physics Department, US Naval Academy} 
\smallskip
\centerline{Annapolis, Maryland}
\smallskip
\centerline{ita@usna.edu} 
  
\bigskip  
                
\begin{abstract}
Using the action for the instanton representation of Plebanski gravity (IRPG), we construct minisuperspace solutions restricted to diagonal variables.  We have treated the Euclidean signature case with zero cosmological constant, depicting a gravitational analogy to free particle motion.  This paper provides a testing ground for the IRPG for a simple case, which will be extended to the full theory in future work.
\end{abstract}
\end{titlepage}

\section{Introduction}

In \cite{EYOITA} a reformulation of General Relativity has been presented, which we refer to as the instanton representation of Plebanski gravity.  The basic variables for the instanton representation of Plebanski are a $SO(3,C)$ gauge connection $A^a_{\mu}$ and a 3 by 3 matrix $\Psi_{ae}$ which takes its values in two copies of $SO(3,C)$.  The instanton representation can be related to the Ashtekar formulation of gravity \cite{ASH2} by a noncanonical transformation.  As shown in \cite{EYOITA}, both formulations of gravity can be seen as sister theories arising from the same mother theory, namely Plebanski's theory of gravity \cite{PLEBANSKI}.  A first order of business in the application of any new formulation is to test that it can produce some known solutions, and the next natural step is to apply the intuition thus gained to the construction of new solutions.  In this paper we will use the instanton representation to construct minisuperspace Euclidean signature solutions, defined on the set of configurations where all variables are spatially homogeneous.\par 
\indent 
The organization of this paper will be as follows.  In the remainder of this section we will provide some background on the instanton representation and its starting action, and then perform a restriction of this action to minisuperspace.  In section 3 we write down the equations of motion for the reduced action.  In section 4, starting from the minisuperspace equations of motion, we carry out a restriction to the diagonal sector, verify the consistency, and solve the resulting equations for zero cosmological constant.  In section 5 we construct the spacetime metric from the instanton representation variables.  The reality conditions on the metric confine our solutions to the Euclidean signature case.  In section 6 we provide a summary of our results and a brief discussion.  

\subsection{The starting action: Reduction to minisuperspace}

The action for the instanton representation of Plebanski gravity is given in 3+1 form by \cite{EYOITA}\footnote{For index conventions we use lower case symbols from the beginning of the Latin alphabet $a,b,c,\dots$ to denote internal $SO(3,C)$ indices, and from the middle $i,j,k,\dots$ to denote spatial indices, each taking values $0-3$.}
\begin{eqnarray}
 \label{THEACTION}
I_{Inst}=\int{dt}\int_{\Sigma}d^3x\Psi_{ae}B^i_e\dot{A}^a_i+A^a_0B^i_eD_i\Psi_{ae}\nonumber\\
+(\hbox{det}B)N^i(B^{-1})^d_i\epsilon_{dae}\Psi_{ae}
-iN(\hbox{det}B)^{1/2}\sqrt{\hbox{det}\Psi}(\Lambda+\hbox{tr}\Psi^{-1}).
\end{eqnarray}
\noindent
The dynamical variables are a $SO(3,C)$ gauge connection $A^a_{\mu}=(A^a_0,A^a_i)$ and a 3 by 3 matrix $\Psi_{ae}\in{SO}(3,C)\otimes{SO}(3,C)$.  The quantity 
\begin{eqnarray} 
\label{MAGNETIC}
B^i_a=\epsilon^{ijk}\partial_jA^a_k+{1 \over 2}\epsilon^{ijk}f_{abc}A^b_jA^c_k
\end{eqnarray} 
\noindent 
is the magnetic field of the spatial connection $A^a_i$, and we require that $(\hbox{det}B)\neq{0}$.  The object $D_i$ is the $SO(3,C)$ covariant derivative, which acts on any 3-vector $v_a\in{SO}(3,C)$ as 
\begin{eqnarray} 
\label{ACTS}
D_iv_a=\partial_iv_a+f_{abc}A^b_iv_c
\end{eqnarray} 
\noindent 
with structure constants $f_{abc}=\epsilon_{abc}$.  The variable $\Psi_{ae}$ can be written as
\begin{eqnarray}
\label{THEVARIABLE}
\Psi^{-1}_{ae}=-{\Lambda \over 3}\delta_{ae}+\psi_{ae},
\end{eqnarray} 
\noindent
where $\psi_{ae}$ is the self-dual part of the Weyl curvature tensor expressed in $SO(3)$ language.  The Petrov type of a spacetime is determined according to the degeneracy of the eigenvalues of $\psi_{ae}$ and the number of linearly independent eigenvectors \cite{PENROSERIND}, \cite{MACCALLUM}.  From (\ref{THEVARIABLE}) we must have $(\hbox{det}\Psi)\neq{0}$, a nondegeneracy condition which limits the equivalence of (\ref{THEACTION}) with General Relativity to spacetimes of Petrov Types I, D and O.\par 
\indent 
The quantities $N$ and $N^i$ are the lapse function and shift vector from metric General Relativity.  The variables $A^a_0$, $N^i$ and $N$ are auxiliary fields whose equations of motion yield the Gauss' law, vector and Hamiltonian constraints
\begin{eqnarray} 
\label{CONSTRAINTS}
B^i_eD_i\Psi_{ae}=0;~~\epsilon_{dae}\Psi_{ae}=0;~~\Lambda+\hbox{tr}\Psi^{-1}=0.
\end{eqnarray} 
\noindent 
To construct a general solution for (\ref{THEACTION}), the constraints (\ref{CONSTRAINTS}) must be solved in conjunction with the equations of motion for the dynamical variables $A^a_i$ and $\Psi_{ae}$.  The disentanglement of coordinate from physical degrees of freedom in solutions to the equations of motion for General Relativity is a difficult problem in part due to the inherent nonlinearity of the Einstein equations.  Due to the nature of the instanton representation variables, on any solution to these equations one has a clean separation of physical from gauge degrees of freedom.  For the purposes of this paper, we will illustrate this feature for the minisuperspace case.
\subsection{Reduction to minisuperspace}
In this paper we will construct spatially homogeneous solutions for the instanton representation for zero cosmological constant.  Rather than construct these solutions directly from the equations of motion for (\ref{THEACTION}), which refers to the full theory, we will use a restriction of (\ref{THEACTION}) spatially homogeneous variables as the starting point.  We will refer to this configuration, where the dynamical variables are spatially homogeneous, as minisuperspace.  Hence the first order of business will be to set all spatial derivatives occuring in (\ref{THEACTION}) to zero, after which we will compute the equations of motion and some solutions.  Some useful relations for minisuperspace as so defined are
\begin{eqnarray}
\label{THEACTION1}
B^i_e=(\hbox{det}A)(A^{-1})^i_e;~~C_{ae}\equiv{A}^a_iB^i_e=\delta_{ae}(\hbox{det}A).
\end{eqnarray}
\noindent
Note that the condition $(\hbox{det}B)\neq{0}$ in minisuperspace is equivalent to the condition that $(\hbox{det}A)\neq{0}$, which will be used throughout this paper.\par
\indent
We will now reduce the action (\ref{THEACTION}) to minisuperspace, reducing each term in turn.  First, the canonical one-form in minisuperspace reduces to  
\begin{eqnarray}
 \label{THEACTION2}
\Psi_{ae}B^i_e\dot{A}^a_i=\Psi_{ae}(\hbox{det}A)(A^{-1})^i_e\dot{A}^a_i\equiv\Psi_{ae}(\hbox{det}A)\dot{X}^{ae},
\end{eqnarray}
\noindent
where we have defined the left invariant one forms (left invariant with respect to the group of time-independent coordinate transformations) $\dot{X}^{ae}$, given by\footnote{This is purely for notational purposes, and is not meant to suggest 
that $\dot{X}^{ae}$ is a total time derivative except for certain special connection configurations.}
\begin{eqnarray}
 \label{THEACTION3}
\dot{X}^{ae}=(A^{-1})^i_e\dot{A}^a_i.
\end{eqnarray}
\noindent
Next, the Gauss` law constraint, which is smeared by the auxiliary field $A^a_0$ in (\ref{THEACTION}), is given by the contraction of $B^i_e$ with $D_i\Psi_{ae}$, the covariant derivative of $\Psi_{ae}$ seen as a $SO(3,C)$ tensor of second rank
\begin{eqnarray}
 \label{THEACTION4}
G_a=B^i_eD_i\Psi_{ae}=B^i_e\bigl(\partial_i\Psi_{ae}+f_{abc}A^b_i\Psi_{ce}+f_{ebc}A^b_i\Psi_{ac}\bigr).
\end{eqnarray}
\noindent
Using (\ref{THEACTION1}) and the spatial homogeneity of $\Psi_{ae}$, then (\ref{THEACTION4}) reduces to
\begin{eqnarray}
 \label{THEACTION5}
G_a=(\hbox{det}A)\bigl(f_{aec}\Psi_{ce}+f_{ebc}\delta_{be}\Psi_{ae}\bigr)=-(\hbox{det}A)f_{ace}\Psi_{ce}
\end{eqnarray}
\noindent
on account of antisymmetry of the structure constants.\par
\indent
The vector constraint, which is smeared by the auxiliary field $N^i$ in (\ref{THEACTION}), upon making use of (\ref{THEACTION1}), simplifies to
\begin{eqnarray}
 \label{THEACTION6}
H_i=(\hbox{det}B)N^i(B^{-1})^d_i\epsilon_{dae}\Psi_{ae}=(\hbox{det}A)N^iA^d_i\epsilon_{dae}\Psi_{ae},
\end{eqnarray}
\noindent
and the Hamiltonian constraint, smeared by the auxiliary field $N$, reduces to
\begin{eqnarray}
 \label{THEACTION7}
H=(\hbox{det}B)^{1/2}\sqrt{\hbox{det}\Psi}(\Lambda+\hbox{tr}\Psi^{-1})=(\hbox{det}A)\sqrt{\hbox{det}\Psi}(\Lambda+\hbox{tr}\Psi^{-1}).
\end{eqnarray}
\noindent
Using all of these results, then the starting action (\ref{THEACTION}) reduced to minisuperspace is given by
\begin{eqnarray}
 \label{THEACTION8}
I_{Inst}=\int{dt}(\hbox{det}A)\Bigl[\Psi_{ae}\dot{X}^{ae}+(N^iA^d_i-A^d_0)\epsilon_{dae}\Psi_{ae}
-iN\sqrt{\hbox{det}\Psi}(\Lambda+\hbox{tr}\Psi^{-1})\Bigr],
\end{eqnarray}
\noindent
where we have omitted an insignificant numerical factor corresponding to the characteristic length scale cubed of 3-space.  In this paper we will consider the case of nondegenerate magnetic fields, where $(\hbox{det}A)\neq{0}$.

\section{Equations of motion}

Using (\ref{THEACTION8}) as the starting action, we can write the equations of motion for the auxiliary fields $N^i$, $A^d_0$ and $N$, given respectively by
\begin{eqnarray}
 \label{THEACTION9}
A^d_i\epsilon_{dae}\Psi_{ae}=0;~~\epsilon_{dae}\Psi_{ae}=0;~~\sqrt{\hbox{det}\Psi}(\Lambda+\hbox{tr}\Psi^{-1})=0,
\end{eqnarray}
\noindent
where we have cancelled off a common factor of $(\hbox{det}A)\neq{0}$.  Equations (\ref{THEACTION9}) are respectively the vector, Gauss' law and Hamiltonian constraints $(H_i,G_a,H)$ in minisuperspace.  Note that Gauss' law and the vector constraints both imply that $\Psi_{ae}=\Psi_{ea}$ must be symmetric.\par
\indent
Having examined the consequences of the initial value constraints, we will now find the equations of motion for the dynamical variables $\Psi_{ae}$ and $A^a_i$.  The Euler--Lagrange equation of motion for $\Psi_{ae}$ from the Lagrangian $L_{Inst}$ in (\ref{THEACTION8}) is given by
\begin{eqnarray}
 \label{THEACTION10}
{{\partial{L}_{Inst}} \over {\partial\Psi_{ae}}}=
\dot{X}^{ae}+f^{aed}(N^iA^d_i-A^d_0)+iN\sqrt{\hbox{det}\Psi}(\Psi^{-1}\Psi^{-1})^{ea}=0,
\end{eqnarray}
\noindent
where we have used $(\hbox{det}A)\neq{0}$ and $\dot{X}^{ae}$ from (\ref{THEACTION3}).\par 
\indent
Next, to obtain the equation of motion for $A^a_i$ we must vary (\ref{THEACTION8}) with respect to $A^a_i$ 
\begin{eqnarray}
 \label{THEACTION11}
{d \over {dt}}\Bigl({{\partial{L}_{Inst}} \over {\partial\dot{A}^b_j}}\Bigr)={{\partial{L}_{Inst}} \over {\partial{A}^b_j}}.
\end{eqnarray}
\noindent
Note that the contributions from the constraint terms of (\ref{THEACTION8}) are directly proportional to the constraints themselves, and therefore will vanish on-shell.  The only term not of this form is the $\Psi_{ae}\dot{X}^{ae}$ term, which will provide a nontrivial contribution to the equations.  So it remains to apply the Leibniz rule to this term to obtain the desired equations.  To keep organized, let us compute the constituents of (\ref{THEACTION11}).  The right hand side of (\ref{THEACTION11}) is given by
\begin{eqnarray}
 \label{THEACTION12}
{\partial \over {\partial{A}^b_j}}(\Psi_{ae}(\hbox{det}A)(A^{-1})^i_e\dot{A}^a_i)\nonumber\\
=\Psi_{ae}(\hbox{det}A)(A^{-1})^j_b(A^{-1})^i_e\dot{A}^a_i
-\Psi_{ae}(\hbox{det}A)(A^{-1})^i_d\Bigl({{\partial{A}^d_k} \over {\partial{A}^b_j}}\Bigr)\nonumber\\
=\Psi_{ae}(\hbox{det}A)(A^{-1})^j_b(A^{-1})^i_e\dot{A}^a_i
-\Psi_{ae}(\hbox{det}A)(A^{-1})^i_b(A^{-1})^j_e\dot{A}^a_i,
\end{eqnarray}
\noindent
and the left hand side upon using the Leibniz rule by
\begin{eqnarray}
 \label{THEACTION13}
{d \over {dt}}(\Psi_{be}(\hbox{det}A)(A^{-1})^j_e)
=\dot{\Psi}_{be}(\hbox{det}A)(A^{-1})^j_e\nonumber\\
+\Psi_{be}(\hbox{det}A)(A^{-1})^m_d\dot{A}^d_m(A^{-1})^j_e
-\Psi_{be}(\hbox{det}A)(A^{-1})^j_d\dot{A}^d_m(A^{-1})^m_e.
\end{eqnarray}
\noindent
Putting the results of (\ref{THEACTION12}) and (\ref{THEACTION13}) into (\ref{THEACTION11}) and dividing by $(\hbox{det}A)\neq{0}$, then we have
\begin{eqnarray}
 \label{THEACTION14}
\Psi_{ae}(A^{-1})^j_b(A^{-1})^i_e\dot{A}^a_i-\Psi_{ae}(A^{-1})^j_e(A^{-1})^i_b\dot{A}^a_i\nonumber\\
=\dot{\Psi}_{be}(A^{-1})^j_e+\Psi_{be}(A^{-1})^m_d\dot{A}^d_m(A^{-1})^j_e-\Psi_{be}(A^{-1})^j_d\dot{A}^d_m(A^{-1})^m_e.
\end{eqnarray}
\noindent
Multiplying (\ref{THEACTION14}) by $A^g_j$ and using the definition (\ref{THEACTION3}), the the equation of motion for $A^a_i$ is equivalent to the following equations
\begin{eqnarray}
 \label{THEACTION15}
\delta_{bg}\Psi_{ae}\dot{X}^{ae}-\Psi_{ag}\dot{X}^{ab}=\dot{\Psi}_{bg}+\Psi_{bg}\dot{T}-\Psi_{be}\dot{X}^{ge},
\end{eqnarray}
\noindent
where we have defined $T=\delta_{ae}X^{ae}$ as the trace.  So the equations of motion for the instanton representation in minisuperspace can be written as
\begin{eqnarray}
\label{THEACTION16}
\dot{X}^{ae}+f^{aed}(-A^d_0+N^iA^d_i)+iN\sqrt{\hbox{det}\Psi}(\Psi^{-1}\Psi^{-1})^{ea}=0;\nonumber\\
\dot{\Psi}_{bg}+\Psi_{bg}\dot{T}-\Psi_{be}\dot{X}^{ge}-\delta_{bg}\Psi_{ae}\dot{X}^{ae}+\Psi_{ag}\dot{X}^{ab}=0.\nonumber\\
\Psi_{ae}=\Psi_{ea};~~\Lambda+{1 \over {\lambda_1}}+{1 \over {\lambda_2}}+{1 \over {\lambda_3}}=0,
\end{eqnarray}
\noindent
where $\lambda_1$, $\lambda_2$ and $\lambda_3$ are the eigenvalues of the symmmetric $\Psi_{ae}$.

\section{Diagonal Ansatz}

\noindent
Having written down the desired equations of motion, we will further simplify our analysis by restriction to the case where all variables are diagonal 
\begin{eqnarray}
\label{THEACTION17}
A^a_i=\delta^a_iA^a_a;~~\Psi_{ae}=\delta_{ae}\Psi_{ee},
\end{eqnarray}
\noindent
with no summation over repeated indices.  This can in one interpretation be regarded as a the choice of a gauge.  Our method for justifying the validity of this choice will be to demonstrate that (\ref{THEACTION17}) is preserved by the equations of motion in congruity with the constraints, both given in (\ref{THEACTION16}).  In addition to verifying this, we will also verify that the off-diagonal parts of the equations of motion are self-consistently zero when the Ansatz (\ref{THEACTION17}) is applied.  We will divide this task into (i) Verifying that the antisymmetric parts of the equations of motion vanish and (ii) verifying that the off-diagonal symmetric parts also vanish.\par 
\indent 
\subsection{Self-consistency of the off-diagonal components of the equations of motion}
We will first find the antisymmetric part of the $\dot{X}^{ae}$ equations of motion by contracting the first equation of (\ref{THEACTION16}) with $f_{aeg}$.  This yields
\begin{eqnarray}
\label{CONSIST}
2N^iA^g_i=-\bigl(\epsilon_{aeg}\dot{X}^{ae}-2A^g_0\bigr),
\end{eqnarray}
\noindent
where the $(\Psi^{-1}\Psi^{-1})^{ae}$ term has cancelled out on account of its being symmetric from the vector constraint.  This enables us to solve for the shift vector
\begin{eqnarray}
\label{CONSIST1}
N^i=-{1 \over 2}\epsilon^{gae}(A^{-1})^i_g\dot{X}^{ae}+A^g_0(A^{-1})^i_g\nonumber\\
=-{1 \over 2}\epsilon^{gae}(A^{-1})^i_g(A^{-1})^j_e\dot{A}^a_j+A^g_0(A^{-1})^i_g.
\end{eqnarray}
\noindent
Since we are restricted to diagonal connections, the first term of (\ref{CONSIST1}) can be written as
\begin{eqnarray}
\label{CONSIST2}
{1 \over 2}\epsilon^{ijk}A^a_k\dot{A}^a_j(\hbox{det}A)^{-1}\longrightarrow{1 \over 2}\epsilon^{iaa}A^a_aA^a_a(\hbox{det}A)^{-1}=0,
\end{eqnarray}
\noindent
on account of antisymmetry of the epsilon symbol.\footnote{The left hand side of (\ref{CONSIST2}) looks like a kind of orbital angular momentum on the space of homogeneous connections.  This term is zero when the connection is diagonal.}  The shift vector then reduces to
\begin{eqnarray}
\label{CONSIST3}
N^i={{A^i_0} \over {A^i_i}}.
\end{eqnarray}
\noindent
So the antisymmetric part of the $\dot{X}^{ae}$ equation of motion is self-consistently zero for diagonal connections provided that the shift vector is given by (\ref{CONSIST3}).\footnote{Equation (\ref{CONSIST3}) features a direct correlation between the gauge degrees of freedom in metric General Relativity, the shift vector $N^i$, and in Yang--Mills theory, the temporal connection components $A^a_0$.  The result is that the spatial and internal indices $i,j,k\rightarrow{a},b,c$ are now on the same footing for these particular degrees of freedom.}\par 
\indent 
The symmetric part of the equations of motion are given by
\begin{eqnarray}
\label{IMPLIESTHATONE}
\dot{X}^{(ae)}+iN\sqrt{\hbox{det}\Psi}(\Psi^{-1}\Psi^{-1})^{ea}=0;\nonumber\\
\dot{\Psi}_{bg}+\Psi_{bg}\dot{T}-\Psi_{(be}\dot{X}^{g)e}-\delta_{bg}\Psi_{ae}\dot{X}^{ae}+\Psi_{a(g}\dot{X}^{ab)}=0.
\end{eqnarray}
\noindent
Note that any $a\neq{e}$ components of the first equation of (\ref{IMPLIESTHATONE}) automatically are zero for diagonal variables.  This is clearly the case for the second term, and for the first term we have that the only nontrivial contribution is given by $(A^{-1})^e_e\dot{A}^a_e=0$ for $a\neq{e}$.  So certainly the off-diagonal symmetric part of the $\dot{X}^{ae}$ is zero.  For the second equation of (\ref{IMPLIESTHATONE}), using the fact that $\Psi_{ae}$ is symmetric from the last equation of (\ref{THEACTION16}), then this implies that
\begin{eqnarray}
\label{IMPLIESTHAT}
-\Psi_{be}\dot{X}^{ge}+\Psi_{ge}\dot{X}^{eb}=(\Psi_{gg}-\Psi_{bb})(A^{-1})^i_b\dot{A}^g_i
=(\Psi_{gg}-\Psi_{bb})(A^b_b)^{-1}\dot{A}^g_b=0
\end{eqnarray}
\noindent
with no summation over $g$ and $b$.  For $g=b$ the first factor is automatically zero and for $g\neq{b}$ the third factor is zero since it has a nondiagonal connection.  So using the fact that the off-diagonal Parts of $\Psi_{ae}$ are zero, then their time derivatives are also zero and they remain zero for all time.  The result is that all off-diagonal components of the equations of motion are zero for diagonal $\Psi_{ae}$ and diagonal $A^a_i$, which means that the diagonal Ansatz (\ref{THEACTION17}) is consistent for these components.  All that remains is to find the diagonal components of the equations of motion. 

\subsection{Diagonal components of the equations of motion}
We have shown that the off-diagonal components of the evolution equations (\ref{THEACTION16}) are zero when diagonal variables (\ref{THEACTION17}) are used.  Moreover, this result is preserved under time evolution.  It remains now to evaluate the diagonal components of these evolution equations.  Only $e=a$ will contribute in the first equation of (\ref{THEACTION16}) and $b=g$ in the second equation.  Thus far we have used the Einstein summation convention, but from now on we will dispense with it to keep clear what is being summed over from what is not.  The diagonal components of (\ref{THEACTION16}) for diagonal variables are
\begin{eqnarray}
 \label{THEACTION18}
\dot{X}^{aa}=-iN\sqrt{\hbox{det}\Psi}(\Psi^{-1}\Psi^{-1})^{aa};\nonumber\\
\sum_{a^{\prime}}\Psi_{a^{\prime}a^{\prime}}\dot{X}^{a^{\prime}a^{\prime}}-\Psi_{gg}\dot{X}^{gg}
=\dot{\Psi}_{gg}+\Psi_{gg}\dot{T}-\Psi_{gg}\dot{X}^{gg}=0.
\end{eqnarray}
\noindent
Note that $g$ is a fixed index while $a^{\prime}$ is a dummy index in the second equation of (\ref{THEACTION18}), whereas $a$ is a fixed index in the first equation.  We will illustrate the case for $g=1$, with the remaining cases to follow by cyclic permutation of indices.  For $g=1$ we have
\begin{eqnarray}
 \label{THEACTION19}
\Psi_{11}\dot{X}^{11}+\Psi_{22}\dot{X}^{22}+\Psi_{33}\dot{X}^{33}=\dot{\Psi}_{11}+\Psi_{11}(\dot{X}^{11}+\dot{X}^{22}+\dot{X}^{33})\nonumber\\
\longrightarrow\dot{\Psi}_{11}+\Psi_{11}(\dot{X}^{22}+\dot{X}^{33})=\Psi_{22}\dot{X}^{22}+\Psi_{33}\dot{X}^{33}.
\end{eqnarray}
\noindent
Defining $X^{aa}\equiv{X}^a$ and $\Psi_{aa}\equiv\lambda_a$\footnote{We have identified the diagonal elements of $\Psi_{ae}$ with its eigenvalues, which is consistent with the fact that the off-diagonal elements are zero.} with all off-diagonal terms zero, then we have the equations
\begin{eqnarray}
 \label{THEACTION20}
\dot{X}^1=u\Bigl({1 \over {\lambda_1}}\Bigr)^2;~~
\dot{X}^2=u\Bigl({1 \over {\lambda_2}}\Bigr)^2;~~
\dot{X}^3=u\Bigl({1 \over {\lambda_3}}\Bigr)^2,
\end{eqnarray}
\noindent
where we have defined
\begin{eqnarray}
 \label{THEACTION21}
u=-iN\sqrt{\lambda_1\lambda_2\lambda_3}.
\end{eqnarray}
\noindent
For the eigenvalues we have the equations
\begin{eqnarray}
 \label{THEACTION22}
\dot{\lambda_1}=(\lambda_2-\lambda_1)\dot{X}^2+(\lambda_3-\lambda_1)\dot{X}^3;\nonumber\\
\dot{\lambda_2}=(\lambda_3-\lambda_2)\dot{X}^3+(\lambda_1-\lambda_2)\dot{X}^1;\nonumber\\
\dot{\lambda_3}=(\lambda_1-\lambda_3)\dot{X}^1+(\lambda_2-\lambda_3)\dot{X}^2.
\end{eqnarray}
\noindent
To find the critical points of the action (\ref{THEACTION8}), we must solve (\ref{THEACTION20}) and (\ref{THEACTION22}) simultaneously.

\subsection{General solution for $\Lambda=0$}

To obtain a general solution to (\ref{THEACTION20}) and (\ref{THEACTION22}), let us first eliminate $X^a$ by substituting (\ref{THEACTION20}) into (\ref{THEACTION22}), obtaining
\begin{eqnarray}
 \label{THEACTION23}
\dot{\lambda}_1=u\Bigl[{{\lambda_2-\lambda_1} \over {(\lambda_2)^2}}+{{\lambda_3-\lambda_1} \over {(\lambda_3)^2}}\Bigr];\nonumber\\
\dot{\lambda}_2=u\Bigl[{{\lambda_3-\lambda_2} \over {(\lambda_3)^2}}+{{\lambda_1-\lambda_2} \over {(\lambda_1)^2}}\Bigr];\nonumber\\
\dot{\lambda}_3=u\Bigl[{{\lambda_1-\lambda_3} \over {(\lambda_1)^2}}+{{\lambda_2-\lambda_3} \over {(\lambda_2)^2}}\Bigr].
\end{eqnarray}
\noindent
Also, summing equations (\ref{THEACTION20}), we have
\begin{eqnarray}
\label{THEACTION24}
\dot{X}^1+\dot{X}^2+\dot{X}^3=\dot{T}=u\Bigl[\Bigl({1 \over {\lambda_1}}\Bigr)^2+\Bigl({1 \over {\lambda_2}}\Bigr)^2+\Bigl({1 \over {\lambda_3}}\Bigr)^2\Bigr].
\end{eqnarray}
\noindent
In what follows we will perform all manipulations on the $\lambda_1$ equation, where the other equations follow by cyclic permutation of indices.  The equation of motion for $\lambda_1$ can be written as
\begin{eqnarray}
 \label{THEACTION26}
\dot{\lambda}_1=u\Bigl[{1 \over {\lambda_2}}+{1 \over {\lambda_3}}
-\lambda_1\Bigl(\Bigl({1 \over {\lambda_1}}\Bigr)^2+\Bigl({1 \over {\lambda_2}}\Bigr)^2
+\Bigl({1 \over {\lambda_3}}\Bigr)^2-\Bigl({1 \over {\lambda_1}}\Bigr)^2\Bigr)\Bigr]\nonumber\\
=u\Bigl[{1 \over {\lambda_1}}+{1 \over {\lambda_2}}+{1 \over {\lambda_3}}
-\Big({1 \over {\lambda_1}}\Bigr)^2-\Bigl({1 \over {\lambda_2}}\Bigr)^2-\Bigl({1 \over {\lambda_3}}\Bigr)^2\Bigr].
\end{eqnarray}
\noindent
Using the Hamiltonian constraint
\begin{eqnarray}
 \label{THEACTION25}
\Lambda+{1 \over {\lambda_1}}+{1 \over {\lambda_2}}+{1 \over {\lambda_3}}=0,
\end{eqnarray}
\noindent
in conjunction with (\ref{THEACTION24}) in (\ref{THEACTION25}), then we get the following equations of motion
\begin{eqnarray}
 \label{THEACTION26}
\dot{\lambda}_1+\lambda_1\dot{T}=\Lambda{u};~~\dot{\lambda}_2+\lambda_2\dot{T}=\Lambda{u};~~\dot{\lambda}_3+\lambda_3\dot{T}=\Lambda{u}.
\end{eqnarray}
\noindent
We could construct some simple solutions, if we could get rid of the inconvenient term $u$ from (\ref{THEACTION21}) in the equations.  One obvious way is to consider the $\Lambda=0$ case as we will do in this paper.  Then (\ref{THEACTION26}) integrates directly to
\begin{eqnarray}
 \label{THEACTION27}
\lambda_1(t)=\alpha{e}^{-T(t)};~~\lambda_2(t)=\beta{e}^{-T(t)};~~\lambda_3(t)=\lambda{e}^{-T(t)},
\end{eqnarray}
\noindent
where $\alpha$, $\beta$ and $\lambda$ are numerical constants of integration satisfying the relation
\begin{eqnarray}
\label{THEACTION28}
{1 \over \alpha}+{1 \over \beta}+{1 \over \lambda}=0.
\end{eqnarray}
\noindent
Note that the Hamiltonian constraint (\ref{THEACTION25}) is preserved for all time for $\Lambda=0$ for the solution (\ref{THEACTION27}), since it is invariant under rescaling of the eigenvalues.  Putting this solution back into (\ref {THEACTION24}) we get the following equation to $T$
\begin{eqnarray} 
\label{THEFOLLOWING}
\dot{T}=-iN\sqrt{\alpha\beta\lambda}\Bigl[\Bigl({1 \over \alpha}\Bigr)^2+\Bigl({1 \over \beta}\Bigr)^2
+\Bigl({1 \over \lambda}\Bigr)^2\Bigr]e^{T/2}.
\end{eqnarray}
Defining
\begin{eqnarray}
 \label{THEACTION30}
\eta=\sqrt{\alpha\beta\lambda}\Bigl[\Bigl({1 \over \alpha}\Bigr)^2+\Bigl({1 \over \beta}\Bigr)^2+\Bigl({1 \over \lambda}\Bigr)^2\Bigr]
\end{eqnarray}
\noindent
as well as the objects
\begin{eqnarray}
 \label{THEACTION31}
\eta^1=\sqrt{\alpha\beta\lambda}\Bigl({1 \over \alpha}\Bigr)^2;~~
\eta^2=\sqrt{\alpha\beta\lambda}\Bigl({1 \over \beta}\Bigr)^2;~~
\eta^3=\sqrt{\alpha\beta\lambda}\Bigl({1 \over \lambda}\Bigr)^2,
\end{eqnarray}
\noindent
then (\ref{THEFOLLOWING}) can be solved firectly for $T$
\begin{eqnarray}
\label{THEACTION29}
\dot{T}=-iN\eta{e}^{T/2}\longrightarrow{e}^{-T/2}=e^{-T_0/2}+{i \over 2}\eta\int^t_0N(t^{\prime})dt^{\prime}.
\end{eqnarray}
\noindent
From (\ref{THEACTION29}) the solutions for $X^a$ automatically follow
\begin{eqnarray}
 \label{THEACTION32}
X^a(t)=X^a(0)+{{\eta^a} \over \eta}T(t).
\end{eqnarray}
\noindent
The motion of $X^a$ is the same as that of a free particle in a three dimensional configuration space with momentum components $\eta^a/\eta$ constrained by a mass shell condition (\ref{THEACTION28}).  The coordinates of this particle evolve with respect to a time $T$.  Note that there is complete freedom in the choice of the lapse function $N(t)$, which provides an equivalence class of such motions.

\section{The spacetime metric}

Having written down the solution for the variables of the instanton representation, we can now construct the spacetime metric solving the Einstein equations for minisuperspace.  Recall from \cite{EYOITA} that the 3-metric is a derived quantity given by the following simple formula 
\begin{eqnarray}
 \label{THEACTION33}
h_{ij}=(\hbox{det}\Psi)(\Psi^{-1}\Psi^{-1})^{ae}(B^{-1})^a_i(B^{-1})^e_j(\hbox{det}B).
\end{eqnarray}
\noindent
Using (\ref{THEACTION1}) for minisuperspace, we have
\begin{eqnarray}
 \label{THEACTION34}
h_{ij}=(\hbox{det}\Psi)(\Psi^{-1}\Psi^{-1})^{ae}A^a_iA^e_j,
\end{eqnarray}
\noindent
which for the case of diagonal variables is given by the following matrix form
\begin{displaymath}
 h_{ij}(t)=(\lambda_1\lambda_2\lambda_3)
\left(\begin{array}{ccc}
(a_1/\lambda_1)^2 & 0 & 0\\
0 & (a_2/\lambda_2)^2 & 0\\
0 & 0 & (a_3/\lambda_3)^2\\
\end{array}\right)
,
\end{displaymath}
\noindent
where $a_a=A^a_a$ are the elements of the (diagonal) connection given by
\begin{eqnarray}
 \label{THEACTION35}
a_1(t)=a_0e^{X^1(t)};~~a_2(t)=a_0e^{X^2(t)};~~a_3(t)=a_0e^{X^3(t)},
\end{eqnarray}
\noindent
and $a_0$ is some numerically constant mass scale.  Note that the eigenvalues $\lambda_a$ also have explicit time dependence from the solution (\ref{THEACTION27}).  Hence the metric can be written explicitly as
\begin{displaymath}
 h_{ij}(t)=\alpha\beta\lambda
\left(\begin{array}{ccc}
(a_1(0)/\alpha)^2(e^T)^{2\eta_1/\eta-1} & 0 & 0\\
0 & (a_2(0)/\beta)^2(e^T)^{2\eta_2/\eta-1} & 0\\
0 & 0 & (a_3(0)/\lambda)^2(e^T)^{2\eta_3/\eta-1}\\
\end{array}\right)
\end{displaymath}
\noindent
where $a_a(0)$ are the initial data for the connection and  
\begin{eqnarray}
\label{THEACTION36}
e^T=\Bigl(e^{-T_0/2}+{i \over 2}\eta\int^t_0N(t^{\prime})dt^{\prime}\Bigr)^{-2}=(\hbox{det}A(t))/(\hbox{det}A(0)).
\end{eqnarray}
\noindent
Reality conditions dictate that the spacetime metric $g_{\mu\nu}$ be real, regardless of the complex nature of the dynamical variables.  The line element for Lorentzian signature spacetimes is given by
\begin{eqnarray}
\label{THEACTION40}
ds^2=g_{\mu\nu}dx^{\mu}dx^{\nu}=-N^2dt^2+h_{ij}(dx^i-N^idt)(dx^j-N^jdt),
\end{eqnarray}
\noindent
which is based on a real lapse function and real shift vector $N,N^i$.  Any complex number with real and imaginary parts $a$ and $b$ can be written as a polar decomposition 
\begin{eqnarray} 
\label{THEREAL}
a+bi=\sqrt{a^2+b^2}e^{i\hbox{tan}^{-1}(b/a)}.
\end{eqnarray}
\noindent 
Since the temporal connection components $A^i_0$ are not constrained by the constraints or equations of motion, then the shift vector $N^i=A^i_i/A^i_0$ can always be made real by choosing the phase angle of $A^i_0$ to exactly cancel that of the corresponding diagonal elements $a_i$.\par 
\indent 
This leaves remaining the reality conditions on the lapse function $N$ and the spatial 3-metric $h_{ij}$.  Since the entire time dependence of $h_{ij}(T)$ is contained in $e^T$, then one possibility is that $e^T$ must be real for all time.  We will assume that the initial data $a_1(0)$, $a_2(0)$ and $a_3(0)$ must chosen so that their ratios to $\alpha$, $\beta$ and $\lambda$ respectively are real, where the latter are constrained by (\ref{THEACTION28}).  This should be possible since there are no other constraints on this initial data $a_a(0)$.\par 
\noindent 
Case (i): For $N$ real the following conditions must be satisfied.  The quantity $\eta$ must be pure imaginary so that $e^T$ in (\ref{THEACTION36}) is real for $e^{T_0/2}$ real.  From (\ref{THEACTION30}), one sees that this can be the case if $\alpha,\beta,\lambda$ are all real and either all negative or two items out of this set are positive with one item negative.  The latter is the only possibility consistent with the Hamiltonian constraint (\ref{THEACTION28}).  This makes $h_{ij}$ negative, which is the same sign as the $N^2$ part of the metric.  Therefore this corresponds to Euclidean signature spacetime.\par 
\noindent   
(ii) Case (ii): For $N$ pure imaginary in (\ref{THEACTION36}) it is possible to have $\eta$ real with $e^{T_0/2}$ real, so that $e^T$ is real.  Since the $N^2$ part of the metric is now positive, then to have a Lorentizan signature metric we would need $h_{ij}$ negative.  This is impossible since we need two items out of the set $\alpha,\beta,\lambda$ to be positive with one item negative, which contradicts the requirement that be $\eta$ real.\par 
\noindent

\section{Summary}

\indent
In this paper we have constructed solutions for the minisuperspace-reduced instanton representation action for gravity, restricted to the Euclidean sector for diagonal variables.  The Gauss' law and vector constraints were automatically satisfied, leaving remaining the Hamiltonian constraint which is a simple algebraic relation.  The clean separation of physical from unphysical degrees of freedom was inherent from the equations of motion via three intermediate results: (i) First, the diagonal degrees of freedom evolved separately from the nondiagonal ones.  Hence the choice of zero for the latter was physically justified, perhaps interpretable as a gauge-fixing choice. (ii) Secondly, the shift vector $N^i$, a coordinate-dependent degree of freedom in metric gravity, correlates to the temporal connection component $A^i_0$, a gauge-dependent degree of freedom in gauge-theory as in (\ref{CONSIST3}). (iii) The lapse function $N$ is also, in metric gravity, a coordinate-dependent degree of freedom, and is freely specifiable (e.g. not fixed by equations of motion alone) in the instanton representation.  However, from (\ref{THEACTION36}) one sees that the spatial metric components $h_{ij}$ can be written in a relational matter such that the lapse function $N$ does not explicitly appear.  Hence the entire time evolution of the theory reduces to an evolution in $T=\hbox{ln}\hbox{det}(A)/a_0^3)$,\footnote{The interpretation is that $T$ can be regarded as a time variable on configuration space, a dynamical degree of freedom with respect to which the remaining variables evovle.} labelled by the initial data $a_a(0)$ and $\alpha,\beta$.  We have constructed solutions of Euclidean signature for $\Lambda\neq{0}$.\par 
\indent  
The Euclidean signature case in General Relativity is important for quantum gravity because it deals with gravitational instantons, which comprise the dominant contribution to the path integral for gravity after a Wick rotation to imaginary time.  The next paper we will treat the $\Lambda\neq{0}$ case, and subsequently the full theory taking full gauge-fixing into account.

\end{document}